# Exotic thermoelectric properties of coronene-cyclobutadienoid graphene nanoribbons

C. Yao[1], Chen Kong[1], H. F. Feng[1], Y. Dong[2,a)], L. Huang[2], X. Zhang[3], Z. X. Song[1,a)], and Zhi-Xin Guo[1,a)]

[1]State Key Laboratory for Mechanical Behavior of Materials and School of Materials Science and Engineering, Xi'an Jiaotong University, Xi'an, Shaanxi, 710049, China.

[2]State Key Laboratory of Oral & MaxillofacialReconstruction and Regeneration , National Clinical ResearchCenter for Oral Diseases , Shaanxi Key Laboratory ofStomatology ,Department of Prosthodontics, School ofStomatology , The Fourth Military Medical University.

[3]College of Mechanical and Materials Engineering, Xi'an University, Xi'an 710065, China.

Authors to whom correspondence should be addressed: dongyanfmmu@qq.com; zhongxiaosong@xjtu.edu.cn; zxguo08@xjtu.edu.cn

**Abstract:**

Thermoelectric materials traditionally incorporate heavy metals to achieve low lattice thermal conductivity. However, elements such as Te, Bi, and Pb are costly and pose environmental hazards. In this study, we introduce a novel design strategy for thermoelectric materials, focusing on room-temperature, light-element, and high-ZT materials such as coronene-cyclobutadienoid graphene nanoribbons (cor$_4$GNRs). This material demonstrates a ZT value exceeding 2.1, attributed to its exceptionally low phonon thermal conductivity resulting from its unique edge structure. Importantly, its electrical conductance and Seebeck coefficient remain relatively high and nearly unaffected by the edge structure. This distinct behavior in phonon and electronic transport properties leads to a remarkably high ZT value. Additionally, we discover that applying strain can significantly reduce phonon thermal conductivity, potentially increasing the ZT value to over 3.0. Our findings provide innovative insights for the design and application of advanced thermoelectric materials.

In recent decades, the ever-growing energy demand has become one of the world's most significant challenges [1,2]. Thermoelectric devices can directly convert waste heat into electricity; thus, they are an effective means to address the global energy crisis [3]. The energy conversion efficiency of thermoelectric materials is typically determined by the dimensionless thermoelectric figure of merit (ZT) as follows:

$$ZT = \frac{PT}{K} = \frac{S^2 GT}{K_e + K_{ph}}, \tag{1}$$

where $P = S^2 G$ is the power factor, $K = K_e + K_{ph}$ is the total thermal conductance, $T$ is the absolute temperature, and $G$, $S$, $K_e$ and $K_{ph}$ are the electrical conductance, Seebeck coefficient, electronic thermal conductance and phonon thermal conductance, respectively.

Since the discovery of the thermoelectric effect, early research has primarily focused on metal alloys and semiconductors. These materials are typically composed of heavy elements or intermetallic compounds with covalent bonds, such as $Bi_2Te_3$, PbTe, SiGe, etc [4-7]. Among them, $Bi_2Te_3$ is a typical low-temperature semiconductor metal alloy thermoelectric material and is currently the most widely used thermoelectric material [8]. Nowadays, many new thermoelectric materials have been discovered, such as oxide thermoelectric materials like $NaCo_2O_4$ [9], $CaMnO_3$ [10], $Cu_2X$ (X = S, Se, Te) [11], $CoSb_3$-based skutterudite [12], clathrate [13], half-Heusler [14], and Zintl phase compounds [15]. Nonetheless, most of the thermoelectric materials with high ZT values contain highly toxic heavy metals, leading to two issues: (1) the heavy metals are harmful to the environment; (2) they are limited in nature, and the low natural abundance of the primary constituent elements of commercially available thermoelectric materials makes them too costly to be popularized on a large scale. Therefore, there is a great need to develop environmentally friendly thermoelectric materials composed of ubiquitous elements.

On the other hand, it has been widely reported that nanomaterials can exhibit exotic thermoelectric properties due to the decoupling of thermal conductivity and electronic conductivity at the nanoscale [16]. For example, the ZT of silicon nanowires (SiNWs) (ZT = 0.6) is 60 times larger than that of bulk silicon (ZT = 0.01) at room temperature [17] due to its low thermal conductivity, which is dozens of times smaller than bulk silicon. Similarly, although graphene has a very small ZT due to its particularly high thermal conductivity (3000-5500 W/mK), graphene nanoribbons (GNRs) can exhibit an unusually large ZT of about 1.0, resulting from the substantially decreased thermal conductivity originating from the edge scattering effect [18]. However, GNRs with ZT > 1.0 have not been achieved due to the high technology preparation requirements. To our knowledge, low-temperature thermoelectric materials with ZT > 1.0, composed of ubiquitous elements and experimentally realizable, are still to be explored.

In this work, based on the first principles calculations, we find that the recently synthesized coronene-cyclobutadienoid GNRs (cor$_4$GNRs) [19] are ideal thermoelectric materials with high ZT at room temperature. We find that the thermal conductivity of cor$_4$GNRs saturates at 12 W/mK, resulting in a length-dependent ZT. Moreover, the electronic conductance and Seebeck coefficient can reach up to 150 μS and 1200 μV/K near the Fermi level. As a result, cor$_4$GNRs have an unusually large ZT of 2.1 at room temperature. Furthermore, by applying uniaxial tension, the ZT value can be increased to 3.0. Our research provides a candidate material composed entirely of light elements for the thermoelectric materials field, which can be synthesized on a large scale. Considering that cor$_4$GNRs are green, non-toxic, and can be synthesized on a large scale, our results demonstrate their great potential applications in low-temperature thermoelectric devices.

The phonon-derived thermal conductivity of the cor$_4$GNRs is investigated using the LAMMPS [20] molecular dynamics software package through non-equilibrium molecular dynamics (NEMD) simulations. The ReaxFF force field is employed to describe the interactions between C-H-O atoms (ReaxFF$^{CHO}$), with parameters adopted from Ref. 27 [21]. In the simulations, the time step is set to 0.2 fs. The conjugate gradient algorithm is used for energy minimization. After optimizing the structure, the system is heated to a target temperature T (300 K) for 100 ps ($5 \times 10^5$ steps) using the Nosé-Hoover thermostat [22]. Then, the heat source region and sink region are coupled with the Langevin thermostat by temperature $T_h$ and $T_c$, respectively [23]. Note that $T_h$ = T + $T_0$ and $T_c$ = T − $T_0$, where $T_0$ is fixed at 10 K, and T corresponds to the average temperature of the system in the non-equilibrium steady state. All results presented in this paper are obtained by averaging over 5 ns after a sufficient equilibration time of 5 ns to establish a non-equilibrium stationary state.

Fig. 1(a) shows the cor$_4$GNRs structure model with a length of 5 nm, where the cross-sectional area is defined as A=W·d, with W and d representing the width and thickness of the cor$_4$GNRs, respectively. The width of cor$_4$GNRs is 11.45 Å. Based on previous studies of GNRs, d is set to 3.40 Å [24,25], which is the van der Waals diameter of a carbon atom [26]. Free boundary conditions are considered in the width and height directions, and the fixed boundary condition is applied in the length direction. Fig. 1(b) shows the temperature distribution of cor$_4$GNRs with a length of 5 nm. One can observe that it is linear and smooth in the middle part of the system. Given the temperature gradient, the phonon thermal conductivity ($\kappa_{ph}$) is determined by

$$k_{ph} = \frac{J}{|\nabla T|} = \frac{J}{\Delta T/L}, \qquad (2)$$

where J denotes heat flux per unit area, $\Delta T = 2T_0$ is the temperature difference between the heat source and heat sink regions, and L is the length of the cor$_4$GNRs. As shown in Fig. 1(c), due to the conservation of total energy in the system, the energy exchange rates of the heat source and heat sink remain essentially the same in the stationary state. Note that the thermal conductance and thermal conductivity have a relationship of $K_{ph} = k_{ph}A/L$, where A is the cross-sectional area of the material and L is its length.

The electronic transport coefficients are simulated using the DFT method combined with the non-equilibrium Green's function (NEGF) formalism, implemented in the Atomistix ToolKit (ATK) 2019 software package [27]. We use a standard model in which the cor$_4$GNRs are separated into a central part that connects with the left and right semi-infinite sections [28]. The electrical and thermal currents through the system are calculated using the Landauer–Büttiker formula by integrating the transmission coefficient $T(E)$ within the energy (E) region [29]:

$$I = \frac{2q}{h} \int dE \, T(E)[f_L(E) - f_R(E)], \qquad (3)$$

$$I_Q = \frac{2}{h} \int dE \, T(E)[f_L(E) - f_R(E)](E - \mu), \qquad (4)$$

where the factor of 2 is for the spin, q is the carrier charge, $T(E)$ is the transmission coefficient of the device, and $f_L$ and $f_R$ are the distribution functions of the left and right reservoirs with chemical potentials $\mu_L$ and $\mu_R$, respectively. In this work, we are interested in the linear response of the system, i.e., we assume that $\Delta \mu = \mu_L - \mu_R$, as well as $\Delta T = T_L - T_R$, are infinitesimally small quantities, making the currents linear in these quantities. Therefore, the chemical potential $\mu$ in the thermal current is defined as the average of the left and right chemical potentials. Furthermore, the thermoelectric response functions can be obtained from the unperturbed ground state properties of the system. In the calculations, the tier 3 basis set is adopted with Hartwigsen-Goedecker-Hutter pseudopotentials, and GGA in the form of the PBE functional is utilized to represent the exchange and correlation interactions. The real-space mesh cutoff is chosen to be 119 Hartree, and the k-point meshes are set as Monkhorst-Pack 1 × 1 × 150 for the device in the Brillouin zone. Moreover, the boundary condition along the transport direction is set to be of Dirichlet type. By using a Taylor expansion in powers of $\Delta V$ and $\Delta T$ in Eqs. (3) and (4), we can derive the thermoelectric transport properties as:

$$G = -\frac{I}{\Delta V}\bigg|_{\Delta T=0} = q^2 L_0, \tag{5}$$

$$S = -\frac{\Delta V}{\Delta T}\bigg|_{I=0} = \frac{L_1}{qTL_0}, \tag{6}$$

$$K_e = \frac{I_Q}{\Delta T}\bigg|_{I=0} = \frac{L_2 - L_1^2/L_0}{T}, \tag{7}$$

where $L_n$ (n = 0,1,2) is the following integral:

$$L_n = \frac{2}{h}\int dE\, T(E)\left(-\frac{\partial f}{\partial E}\right)(E-\mu)^n. \tag{8}$$

where $G$, $S$, $K_e$ and $\mu$ are the electronic conductance, Seeback coefficient, electronic thermal conductance, and chemical potential, respectively. Note that the approach employed is based on the study by Esfarjani et al [30]. The derivative of the Fermi function, $\partial f/\partial E$, is referred to as T(E), which is ascertained using the NEGF formalism on the Hamiltonian and the overlap matrices.

We initially investigate the influence of length on the phonon thermal conductivity ($k_{ph}$) of cor$_4$GNRs at room temperature for length (L) ranging from 5 nm to 40 nm. As depicted in Fig. 2, $k_{ph}$ increases slightly with length, adhering to a power law $k_{ph} \propto aL^\beta$ ($\beta = 0.414$), akin to other one-dimensional (1D) materials [31-35]. This behavior highlights the nature of ballistic phonon transport in cor$_4$GNRs, which is particularly pronounced in 1D systems [36]. Despite the significant $\beta$, the $k_{ph}$ of cor$_4$GNRs remains relatively low, rising from 4.98 W/mK to 11.91 W/mK as L varies from 5 nm to 40 nm due to the small coefficient $a$ (2.6) in the $k_{ph}$-L relationship. To validate this, we also computed the $k_{ph}$ of Zigzag-GNRs (ZGNRs) and Armchair-GNRs (AGNRs) with similar widths and the same ReaxFF potential as cor$_4$GNRs in the NEMD simulations. As illustrated in Fig. 2, the $k_{ph}$ of both ZGNRs and AGNRs (referred to as conventional GNRs) is significantly higher than that of cor$_4$GNRs. Specifically, with L = 40 nm, the $k_{ph}$ of ZGNRs and AGNRs reaches up to 280 W/mK and 160 W/mK, respectively, which are 23 and 13 times larger than that of cor$_4$GNRs. This markedly lower $k_{ph}$ in cor$_4$GNRs suggests superior thermoelectric properties compared to conventional GNRs. The variation of $k_{ph}$ with L in ZGNRs and AGNRs also follows the power law $k_{ph} \propto aL^\beta$, but the coefficients $a$ and $\beta$ are considerably larger for ZGNRs ($a$ = 40, $\beta$ = 0.520) and AGNRs ($a$ = 18, $\beta$ = 0.590) than for cor$_4$GNRs. This result indicates that the thermoelectric superiority of cor$_4$GNRs over conventional GNRs becomes more pronounced with increasing L, suggesting the potential for achieving a high ZT in cor$_4$GNRs at room temperature. This extremely low thermal conductivity in cor$_4$GNRs stems from their unique 1D structure. As depicted in Fig. 1, this structure consists of carbon clusters (saturated by hydrogen and oxygen atoms at the edges) connected via a ring of four carbon atoms. Consequently, the edge effect in cor$_4$GNRs induces significant phonon scattering, thereby reducing thermal conductivity [37].

To investigate the electronic transport properties, we employed the NEGF method to calculate the transmission function T(E) for cor$_4$GNRs. As shown in Fig. 3(a), the transmission spectra of cor$_4$GNRs display asymmetry around the Fermi level (E = 0). For both n-type (E > 0) and p-type (E < 0) doping, the number of the first conducting channel remains unity, determined by counting the energy bands at specific energy levels [28]. The absence of the transmission function near the Fermi level is consistent with cor$_4$GNRs behaving as semiconductors with a bandgap of 0.8 eV. By integrating the calculated transmission function T(E), we derive the Seebeck coefficient (S), electrical conductance (G), and electronic thermal conductance ($K_e$) [38]. Figures 3(b)-(d) illustrate these transport coefficients as functions of chemical potential μ at 300 K. Notably, $K_e$ exhibits high values in the energy ranges of [-2.0, -0.5], [0.5, 1.2], and [1.7, 2.0] eV, while phonon thermal conductivity dominates other energy ranges [Fig. 3(b)]. As shown in Fig. 3(c), the electrical conductance also has significant values within the energy ranges of [-2.0, -0.5], [0.5, 1.2], and [1.7, 2.0] eV. However, the Seebeck coefficient of cor$_4$GNRs is substantial in the energy ranges of [-2.0, -1.5], [-0.5, 0.5], and [1.0, 1.5] eV for the chemical potential, with a maximum value of 1200 μV/K at μ = -0.05 eV [Fig. 3(d)]. Since ZT is proportional to the product of $S^2G$, this feature indicates a nontrivial chemical potential-dependent ZT property of cor$_4$GNRs. Consequently, according to Eq. (1), the chemical potential dependence of ZT values is mainly influenced by the Seebeck coefficient and electrical conductance. Therefore, a large ZT value can be achieved by selecting the appropriate chemical potential where both substantial S and G exist.

After calculating all the transport coefficients, we evaluated the thermoelectric performance of cor$_4$GNRs. Figure 4(a) illustrates the chemical potential-dependent ZT of cor$_4$GNRs at room temperature for different lengths. A common feature is the appearance of remarkably high ZT values at chemical potential values where both substantial S and G exist, consistent with our earlier analysis. Moreover, the ZT value increases monotonically with length (L). As L increases up to 40 nm, the maximum ZT value reaches 2.08 at μ = 1.21 eV. It is noteworthy that the length-dependent ZT of cor$_4$GNRs can be further modelled using an exponential formula [39],

$$ZT = ZT_\infty \left[1 - \exp\left(-\frac{L}{L_c}\right)\right], \qquad (9)$$

where $ZT_\infty$ is the fitted fully convergent ZT value, L is the sample length, and L$_c$ is the fitted length describing the transition. According to the fitting results, the converged ZT values are 1.50, 1.43, and 2.10 for μ = -0.35, 0.37, and 1.21 eV, respectively [Fig. 4(b)]. This finding indicates that the ZT of cor$_4$GNRs is not only significantly higher than that of conventional GNRs [40-42] but also comparable to traditional room-temperature thermoelectric materials [43], demonstrating the significant potential applications of cor$_4$GNRs in thermoelectrics.

Finally, we demonstrate that the ZT of cor$_4$GNRs can be effectively enhanced through tensile strain. Figure 5(a) presents the chemical potential-dependent electronic conductance (G). It is observed that tensile strain primarily results in a slight enlargement of the band gap with minimal impact on the magnitude of G. The minor change in G indicates the negligible effect of tensile strain on electronic properties. This observation is corroborated by the calculated chemical potential-dependent Seebeck coefficient, which directly relates to the electronic band structures. As shown in Fig. 5(b), a 5% tensile strain mainly causes a small shift in the peak positions of S. This result suggests that the variation in ZT would be predominantly influenced by the strain-dependent thermal conductivity. The length-dependent thermal conductivity at room temperature is further illustrated in Fig. 5(c), which also follows a power law $k_{ph} \propto aL^\beta$. Notably, the thermal conductivity of cor$_4$GNRs for different lengths significantly decreases under a 5% tensile strain, mainly due to a substantial reduction in the coefficient $a$, which decreases from 2.6 in the unstretched condition to 1.5 in the stretched condition. Additionally, the $\beta$ value in the power law decreases from 0.414 to 0.408, indicating that the thermoelectric performance improves significantly with increasing length. The calculated chemical potential-dependent ZT under a 5% tensile strain is further depicted in Fig. 5(d). Compared to the unstretched condition, a common feature is that the ZT peaks slightly shift to higher and lower energy levels for μ > 0 and μ < 0, respectively, due to the increased band gap leading to a similar shift in G.

It is also observed that the ZT value significantly increases in stretched cor$_4$GNRs. Specifically, around the Fermi level, the previously substantial ZT values of 1.45 and 1.41 for μ = -0.35 and 0.37 without strain increase to 2.19 and 2.19 for μ = -0.51 and 0.51 with L = 40 nm under 5% strain. The overall increase in ZT value is primarily attributed to the decrease in $k_{ph}$ under 5% tensile strain. The highest ZT value is noted at μ = -1.63, reaching 2.92 for L = 40 nm. This value is 1.4 times greater than that without stretching, due to the significant reduction in electronic thermal conductance at μ = -1.63, as shown in Fig. 6(a). Furthermore, the ZT value of stretched cor$_4$GNRs with infinite length can be extrapolated using the method described in Eq. (9). As shown in Fig. 6(b), the ZT values of infinitely long cor$_4$GNRs for μ = -0.51, 0.51, and -1.63 eV reach up to 2.21, 2.27, and 3.02, respectively, after stretching, significantly higher than those of traditional thermoelectric materials such as Bi$_2$Te$_3$ and PbTe [44-46].

In summary, based on the first-principles and molecular-dynamics methods, we studied the thermal and electrical transport properties of an all-light element nanobelt, cor$_4$GNRs. Its room temperature ZT value exceeds 2.1, attributed to its very low thermal conductivity caused by its unique edge structure, while its electrical and thermoelectric coefficients remain relatively high due to minimal edge structure influence. This effective separation of thermoelectric transmission characteristics leads to a high ZT value. Additionally, we found that the application of strain can significantly reduce phonon thermal conductivity, further increasing the ZT value to exceed 3.0. Our research identifies cor$_4$GNRs as a promising candidate material for thermoelectric applications, synthesizable on a large scale.

We acknowledge financial support from the Natural Science Foundation of China (Grant No. 12074301), Science Fund for Distinguished Young Scholars of Shaanxi Province (No. 2024JC-JCQN-09), and the Natural Science Foundation of Shaanxi Province (Grant No. 2023-JC-QN-0768).

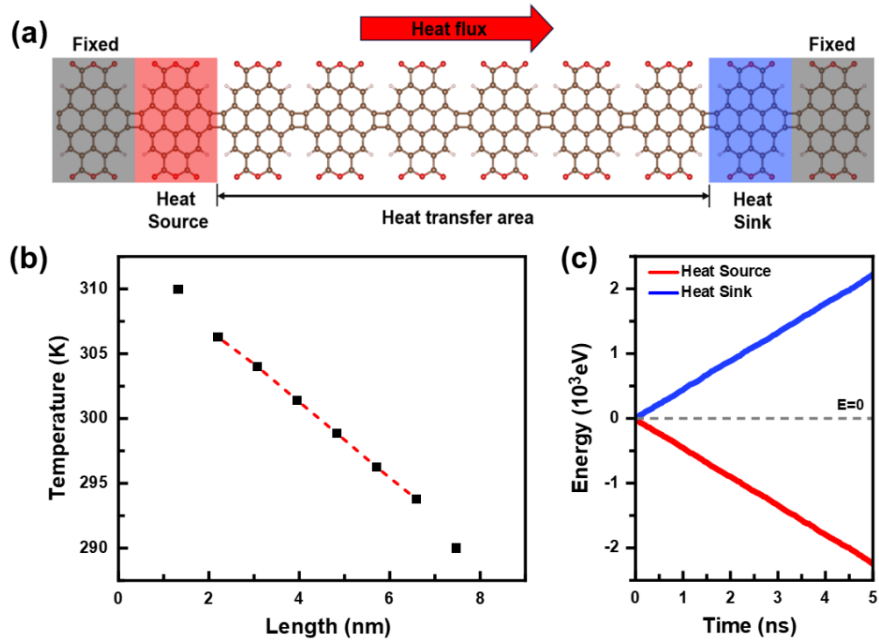

Fig.1 (a) NEMD simulation model of cor$_4$GNR with a length of 5 nm. (b) Temperature profile of cor$_4$GNR after heat equilibration. (c) Cumulative energy of heat source and heat sink under Langevin thermostat as a function of time.

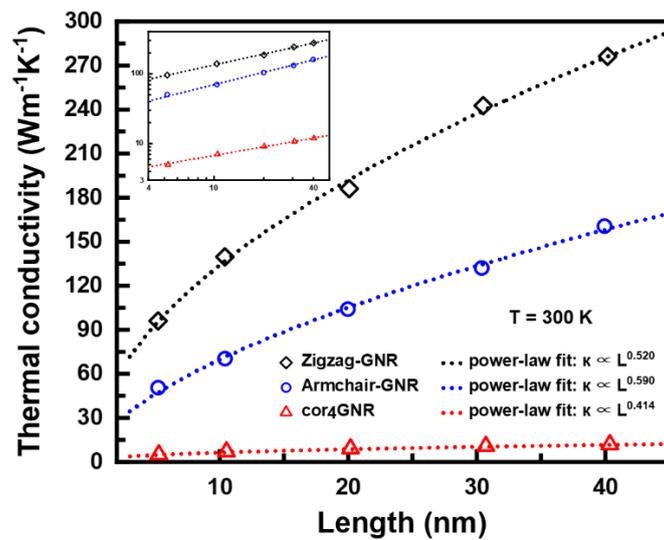

Fig.2 Length dependence of thermal conductivity of cor$_4$GNR with a width of 11.45 Å, armchair-GNR (AGNR) with a width of 8.49 Å and zigzag-GNR (ZGNR) with a width of 11.31 Å at room temperature (300 K), respectively. The inset shows a log-log profile.

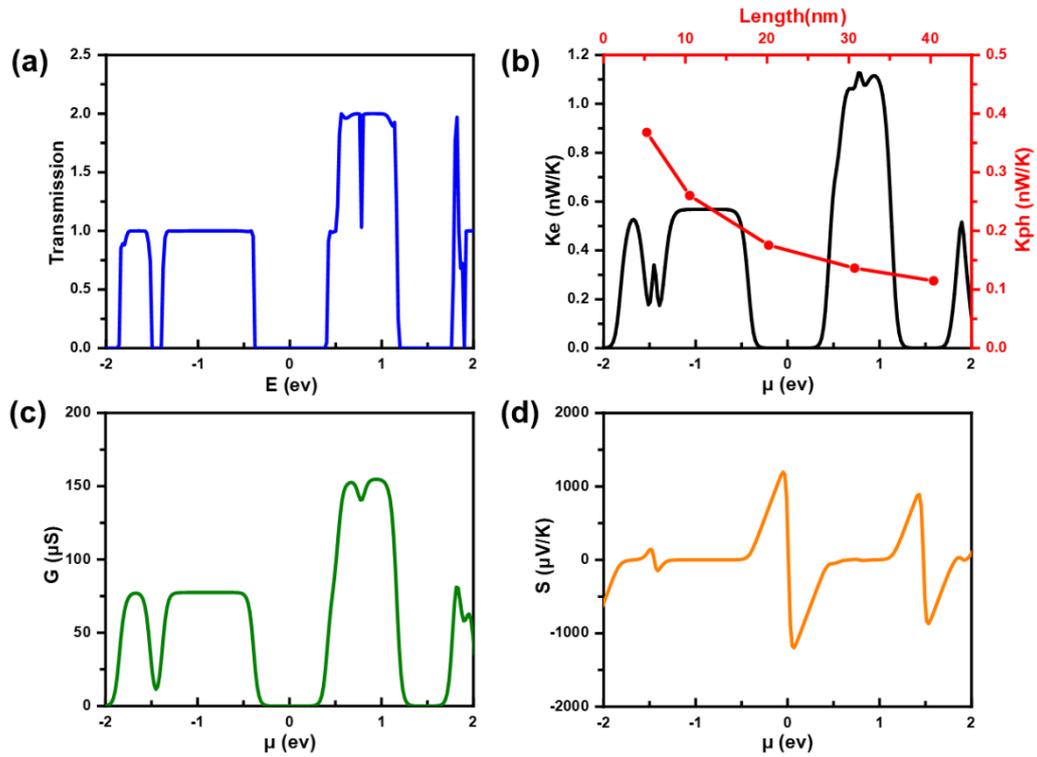

Fig.3 Calculated transport properties and Seebeck coefficient of cor$_4$GNR as a function of chemical potential. (a) Electronic transmission function, (b) Electronic thermal conductance (The red curve shows phonon thermal conductance as a function of length), (c) electrical conductance, and (d) Seebeck coefficient.

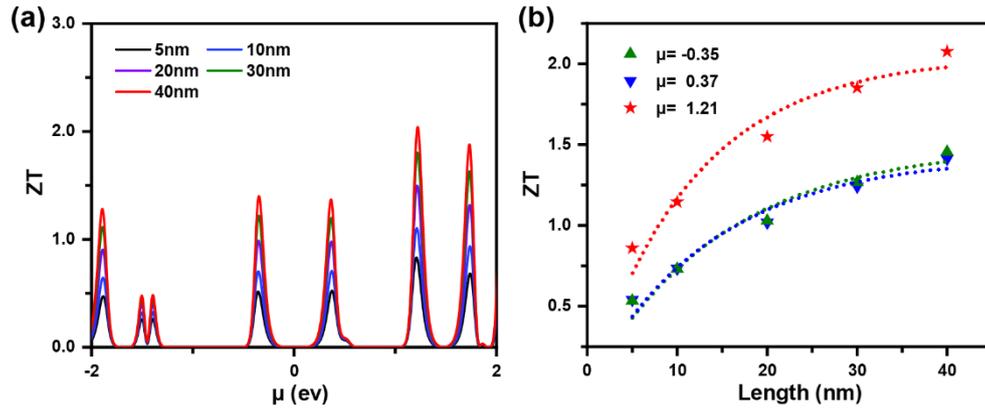

Fig.4 (a) Thermoelectric value (ZT) for cor$_4$GNRs at different lengths as a function of chemical potential. (b) Fitted length dependent ZT for cor$_4$GNRs at various chemical potential, using formular $ZT = ZT_\infty \left[1 - \exp\left(-\frac{L}{L_c}\right)\right]$.

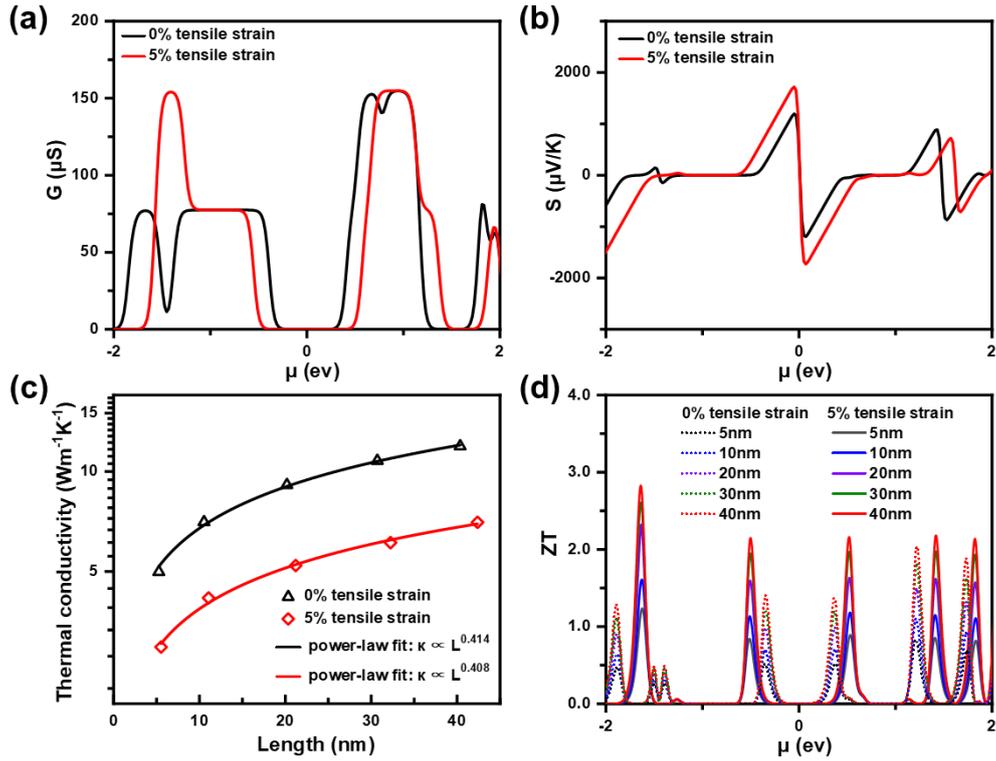

Fig.5 (a) electrical conductance and (b) Seebeck coefficient as a function of chemical potential, (c) length dependent thermal conductivity and (d) chemical potential dependent, ZT before and after applying 5% uniaxial stretching.

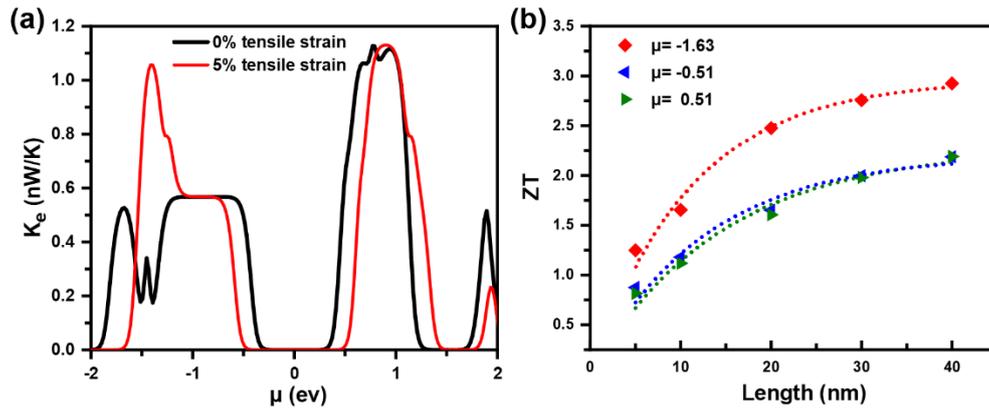

Fig.6 Variation of (a) electrical thermal conductance with chemical potential before and after applying 5% uniaxial stretching. (b) Fitted length dependent ZT for cor$_4$GNRs at various chemical potential after 5% stretching.